\documentclass[onecolumn,aps,superscriptaddress,pre]{revtex4}

\usepackage{amsmath,amssymb,graphicx}
\usepackage{algorithmic}
\usepackage{enumerate}
\usepackage{times}
\usepackage{color}
\usepackage{soul}
\definecolor{yblue}{rgb}{0.06, 0.3, 0.57}
\usepackage[pdftex]{hyperref}
\hypersetup{colorlinks=true,linkcolor=blue,citecolor=blue,urlcolor=blue}

\begin{document}

%\title{Solving the one-dimensional Ising model using mathematical induction: a physically intuitive approach}
\title{Solving the one-dimensional Ising chain via mathematical induction: An intuitive approach to the transfer matrix}

\author{Wenlong Wang}
\email{wenlongcmp@gmail.com}
\affiliation{Department of Physics, Royal Institute of Technology, Stockholm, SE-106 91, Sweden}

\author{Rogelio D\'iaz-M\'endez}
\affiliation{Department of Physics, Royal Institute of Technology, Stockholm, SE-106 91, Sweden}

\author{Raudys Capdevila}
\affiliation{Department of Applied Mathematics, Polytechnical University of Valencia, 46022 Valencia, Spain}

%\author{Mats Wallin}
%\affiliation{Department of Theoretical Physics, Royal institute of technology, Stockholm, SE-106 91, Sweden}
%
%\author{Jack Lidmar}
%\affiliation{Department of Theoretical Physics, Royal institute of technology, Stockholm, SE-106 91, Sweden}
%
%\author{Egor Babaev}
%\affiliation{Department of Theoretical Physics, Royal institute of technology, Stockholm, SE-106 91, Sweden}

\begin{abstract}
The aim of this work is to present a formulation to solve the one-dimensional Ising model using the elementary technique of mathematical induction.
This formulation is physically clear and leads to the same partition function form as the transfer matrix method, which is a common subject in the introductory courses of statistical mechanics.
In this way our formulation is a useful tool to complement the traditional more abstract transfer matrix method. The method can be straightforwardly generalized to other short-range chains, coupled chains and is also computationally friendly. These two approaches provide a more complete understanding of the system, and therefore our work can be of broad interest for undergraduate teaching in statistical mechanics.
\end{abstract}

%\pacs{75.50.Lk, 75.40.Mg, 05.50.+q, 64.60.-i}
\maketitle

\section{Introduction}
The one-dimensional (1D) Ising model~\cite{ising25,brush67} is of fundamental importance in an introductory course on statistical mechanics, because it is connected to many interesting physical concepts; see e.g. Ref.~\cite{Taroni15} for a recent entertaining introduction. %It is one of the simplest nontrivial many-body systems, and is ideal for introducing
Despite the absence of a genuine phase transition, the 1D Ising model still plays a central role in the comprehension of many principles, phenomena, and numerical methods in statistical physics. 
Indeed, the model itself and its large number of variations have become the most common conceptual platform to which discussions are constantly referred to.
Consequently, in the context of teaching, the development of an intuitive comprehension of the many features of the Ising and related models is an instrumental tool for the proper acquisition of the subsequent content.           
Moreover, while the exact solution of the 2D version of the model is a very challenging exercise~\cite{onsager44,Kac52,Kano53,Domb60}, the one-dimensional calculation typically becomes for the students the first contact with an exactly solvable many body model at their reach.
In this way facing a decades-old scientific problem which, with slight modifications, can easily turn on current interest~\cite{kofinger10,sarkanych18,HM08,bach19}. 
There exist several analytical methods to solve the 1D Ising model, some of them providing novel approaches and  interesting view points~\cite{Baxter82,Seth17}.
However, the transfer matrix method is by far the most extended technique in undergraduate lectures, due in part to its wide general use across many physical subjects~\cite{moreno05,pujol07,pujol14,Zad17,Torrico18}. 
The method is simple to use and can be flexibly generalized to cover other interesting models as next-nearest neighbours~\cite{kassan01,baohua02}, Potts spins~\cite{shrock97,shrock10} and coupled chains~\cite{yurishchev01,guidi15}.
%In 1D this method have successfully proven to be quite easy to follow for beginners.
 
%However, due to the lack of axiomatic treatments or alternative formulations. 

There is, though, a common conceptual barrier from novice students to properly understand the basis of the transfer matrix method, unless the students are comfortable with the bond representation.
The latter is unfortunately not always the case as this is usually their first encounter with the Ising model. 
In this line, many teachers find the physical basis of the regular transfer matrix method a bit uneasy to define and explain.
In many cases, after getting familiar with the method, the lack of an axiomatic presentation of the matrix or even an intuitive analogy makes it to be generally seen as a lucky trick whose underlying idea has no link with the physical problem.
%got an idea to solve the 1d Ising model. 
%In this line, many teachers find the physical basis of the regular transfer matrix method a bit uneasy to define and explain.
It is expectable that efforts in the direction of offering physical connections to this trick can be very welcomed in the wide community of statistical physics students and teachers.
In this paper we show that, by using elementary mathematical induction, it is possible to solve the 1D Ising model using simple mathematics where mathematical reduction can be seen from the very first step.
More importantly, we find a working transformation from which the transfer matrix result can be derived.
The advantage of this formulation is that: (1) it is very physically clear and mathematically simple, and (2) it connects the transfer matrix method with an intuitively logical mechanism.
We believe this derivation could be very useful for developing the intuitive comprehension of this topic in introductory lectures of statistical mechanics.

\section{The Transfer Matrix Method}

We start by a quick review of the regular transfer matrix method. 
In the presence of an external field, the 1D Ising Hamiltonian takes the form
\begin{eqnarray}
H = - J \sum_{i=1}^{N} S_i S_{i+1} - h\sum_{i=1}^{N} S_i,
\end{eqnarray}
where $N$ spins $S_i=\pm1$ are placed in a chain, interacting only with their nearest neighbours and with the external field $h$.  
Periodic boundary conditions are imposed by setting $S_{N+1}\equiv S_1$.

The canonical partition function for this system can be expressed as
\begin{eqnarray}
Z_N= \sum_{\{S_i\}} e^{-\beta H(S_1,S_2,\ldots,S_N)} =\sum_{\{S_i\}} e^{\beta\big[ \sum_{i=1}^{N}JS_iS_{i+1} + \frac{1}{2}h\sum_{i=1}^N(S_i+S_{i+1})\big]},
\end{eqnarray}
where $\{S_i\}$ means that the sum runs over all ($2^N$) possible combination of the values of the $N$ variables $S_i$, i.e., over all possible states of the system.
The next step is to use the bond representation and factor the Boltzmann weights to pairwise factors, a decomposition of the bonds
\begin{eqnarray}
Z_N=\sum_{S_1}\sum_{S_2}\cdots\sum_{S_N} \prod_{i=1}^{N} 
e^{\beta\big[ JS_iS_{i+1} + \frac{1}{2}h(S_i+S_{i+1})\big]}.
\end{eqnarray}
Introducing a matrix, the transfer matrix, with the corresponding matrix elements from the factors defined as 
\begin{eqnarray}
P_{S_iS_j}=e^{\beta J S_i S_j +\frac{1}{2} \beta h (S_i+S_j)},
\end{eqnarray}
the partition function can be written as
\begin{eqnarray}
Z_N&=&\sum_{S_1}\sum_{S_2}\cdots\sum_{S_N}  
P_{S_1S_2}P_{S_2S_3}\cdots P_{S_NS_1} \nonumber \\
&=&\sum_{S_1} (P^N)_{S_1S_1} = Tr(P^N).
\label{tmres}
\end{eqnarray}
It is important to realize that there is no matrix in the partition function itself, only matrix elements. 
The summations remarkably happen to resemble matrix multiplications. 
The approach is not very intuitive, mostly because it is difficult to see a priori that this reduction will take place, at least for the first time, and the students have to take the fact that it works.
As we will demonstrate in the next section, this same result can be obtained by means of a physically intuitive formalism, where reduction can be seen in the first place.

Before ending this section, we summarize some important results here for reference. The matrix $P$ is a $2\times2$ matrix as the variables $S_i$ has two possible values. Ordering the spins in the state space as $1$ and $-1$, the matrix reads 
\begin{eqnarray}
P = 
\begin{pmatrix}
e^{\beta (J + h)} & e^{-\beta J} \\
e^{-\beta J} & e^{\beta (J - h)}
\end{pmatrix}
\label{tm}
\end{eqnarray}
 
%If this matrix $P$ can be diagonalized, its trace can be written in the form
%\begin{eqnarray}
%\mathrm{Tr}(P^N)&=&\mathrm{Tr}(RDR^{-1}RDR^{-1}\cdots RDR^{-1})
%\end{eqnarray}
%and applying the cyclic property of the traces it can be reduced to
%\begin{eqnarray}
%\mathrm{Tr}(P^N)&=&\mathrm{Tr}(DR^{-1}RDR^{-1}\cdots RD)\\
%&=&\mathrm{Tr}(D^2R^{-1}RDR^{-1}\cdots R)\\
%&=&\ldots\\
%&=&\mathrm{Tr}(D^N)
%\end{eqnarray}
The partition function can be calculated as
\begin{eqnarray}
Z_N=Tr(P^N)&=&\lambda_1^N+\lambda_2^N,
\label{zll}
\end{eqnarray}
where $\lambda_1$ and $\lambda_2$ are the eigenvalues of the matrix $P$, satisfying the usual characteristic equation 
\begin{eqnarray}
\lambda^2-Tr(P)\lambda+Det(P)=0
\end{eqnarray}
with solutions
\begin{eqnarray}
\lambda_{1,2}=e^{\beta J} \left[ \mathrm{cosh}(\beta h)\pm\sqrt{\mathrm{sinh}^2(\beta h)+e^{-4\beta J}} \right],
\end{eqnarray}
where $\lambda_1>\lambda_2$ for any value of $h$.
Consequently, in the thermodynamic limit, the term involving $\lambda_2$ in the right hand side of Eq.~(\ref{zll}) can be neglected  
\begin{eqnarray}
&Z_N&\xrightarrow[N \to \infty]{} \lambda_1^N
\end{eqnarray}
ending up with a partition function in the form
\begin{eqnarray}
&Z&\approx \lambda_1^N = e^{N \beta J} \left[ \mathrm{cosh}(\beta h)+\sqrt{\mathrm{sinh}^2(\beta h)+e^{-4\beta J}} \right]^N.
\label{finalZ}
\end{eqnarray}

Importantly, the free energy $f$ and the magnetization $m$ per spin are
\begin{eqnarray}
f(h,T)&=&-\frac{1}{N}k_BT\ln(Z)=-k_BT\ln\lambda_1 \nonumber \\
&=&-J -k_BT \ln \left[\mathrm{cosh}(\beta h)+\sqrt{\mathrm{sinh}^2(\beta h)+e^{-4\beta J}} \right], \\
m(h,T)&=&-\frac{\partial f}{\partial h} \nonumber \\
&=&\frac{\sinh(\beta h)}{\sqrt{\sinh^2(\beta h)+e^{-4\beta J}}},
\end{eqnarray}
from which,
since $\sinh(\beta h)\xrightarrow[\beta h \to 0]{} 0 $,
it is clear that no spontaneous magnetization occurs at any finite temperature in the absence of an external field.

\section{Solving via Mathematical Induction}

We now consider the more intuitive approach expressed in the following question: given the partition function for $N$ spins $Z_N$, could one construct the partition function for $N+1$ spins $Z_{N+1}$?
In this situation, the new partition function $Z_{N+1}$ seems likely to be expressible in terms of the old partition function $Z_N$, after all, the new spin only change the interactions locally.
Without loss of generality, we assume this extra spin $S_{N+1}$ is inserted at the right end of the chain, i.e., interacting with spins $S_N$ and $S_1$.

\begin{figure}
	\includegraphics[width=.49\columnwidth]{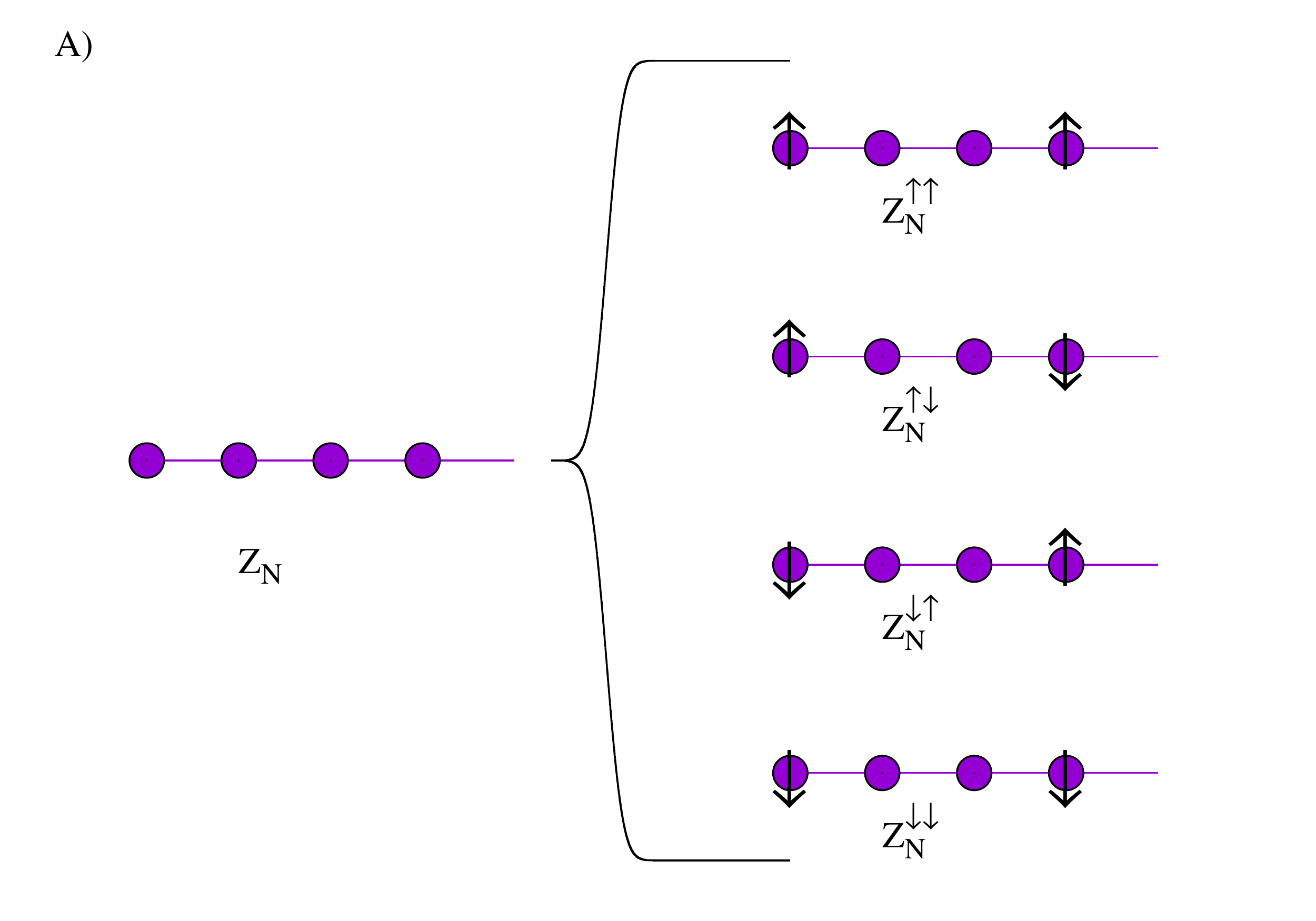}	
	\includegraphics[width=.49\columnwidth]{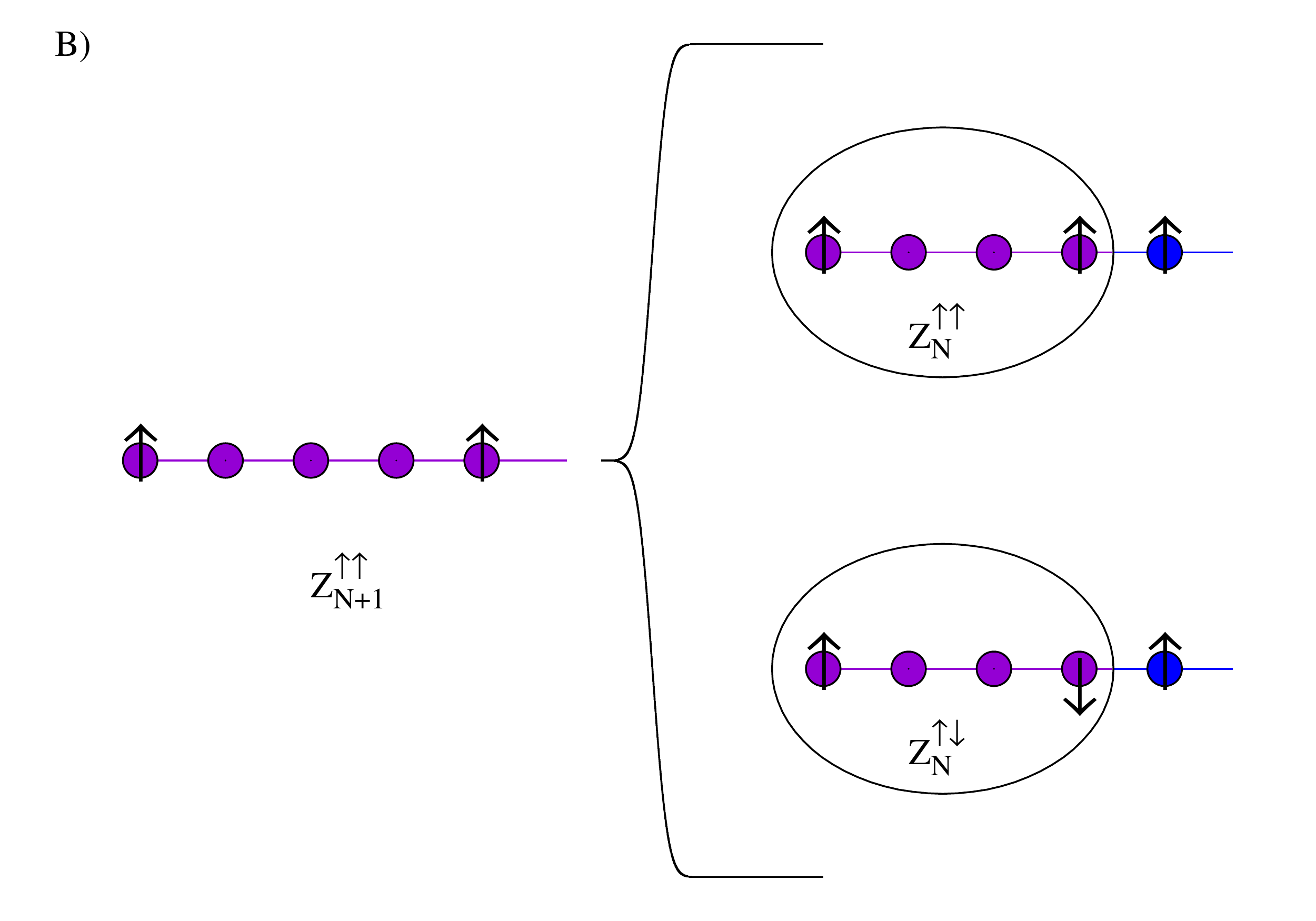}	
	\caption{
		Schematic representation of the configuration space. 
		Spins (circles) with no explicit orientation can take any of the two values $\pm1$, and extra bonds at the right of the chain indicates periodic interactions.
		Panel A: the configuration space of a chain of $N=4$ spins is decomposed in four sets of configurations by considering the four possible orientations of the spins at the extremes.
		Panel B: for a chain of $N+1=5$ spins, the subset $(\uparrow \uparrow)$ is the union of the subsets $(\uparrow\uparrow)$ and $(\uparrow\downarrow)$ of the chain with $N=4$ spins, plus an extra spin $S_{N+1}=1$. 
	}	
	\label{fig}
\end{figure}

In order to keep track of the modifications raised by the new spin, we now split the configuration space in four subsets, according to the four possible orientations of the first and the last spins. This is a preparation step for mathematical induction, as we know one can adjust the Boltzmann factors properly when a new spin is added if the additional interactions (with $S_1$ and $S_N$) are fully known.
All configurations with $S_1=S_N=1$ form the subset $\{\uparrow\uparrow\}$, configurations with $S_1=1$ and $S_N=-1$ form the subset $\{\uparrow\downarrow\}$, and so on, each of them has $2^{N-2}$ configurations.
That is to say
\begin{eqnarray}
\{S_i\}=\{\uparrow\uparrow\}\cup\{\uparrow\downarrow\}\cup\{\downarrow\uparrow\}\cup\{\downarrow\downarrow\}.
\end{eqnarray} 
Such a splitting is illustrated in Fig.~\ref{fig}~A.
Extending the split to the sum in the partition function, one can define four terms associated to these sectors $Z_N^{\uparrow \uparrow}$, $Z_N^{\uparrow \downarrow}$, $Z_N^{\downarrow \uparrow}$ and $Z_N^{\downarrow \downarrow}$, 
such that 
\begin{eqnarray}
\label{znpp}
Z_N^{\uparrow\uparrow}&=&\sum_{\{\uparrow\uparrow\}} e^{-\beta H(S_1,S_2,\ldots,S_N)}
=\sum_{\{S_i\} : S_1=S_N=1} e^{-\beta H(S_1,S_2,\ldots,S_N)}\\
Z_N^{\uparrow\downarrow}&=&\sum_{\{\uparrow\downarrow\}} e^{-\beta H(S_1,S_2,\ldots,S_N)}
=\sum_{\{S_i\} : S_1=1,S_N=-1} e^{-\beta H(S_1,S_2,\ldots,S_N)}\\
Z_N^{\downarrow\uparrow}&=&\sum_{\{\downarrow\uparrow\}} e^{-\beta H(S_1,S_2,\ldots,S_N)}
=\sum_{\{S_i\} : S_1=-1,S_N=1} e^{-\beta H(S_1,S_2,\ldots,S_N)}\\
Z_N^{\downarrow\downarrow}&=&\sum_{\{\downarrow\downarrow\}} e^{-\beta H(S_1,S_2,\ldots,S_N)}
=\sum_{\{S_i\} : S_1=S_N=-1} e^{-\beta H(S_1,S_2,\ldots,S_N)}
\end{eqnarray}
and the partition function can be expressed as
\begin{eqnarray}
\label{terms}
Z_N = Z_N^{\uparrow \uparrow}+
Z_N^{\uparrow \downarrow}+
Z_N^{\downarrow \uparrow}+
Z_N^{\downarrow \downarrow}.
\end{eqnarray}

The next step is to explicitly find the transformations of each component of the partition function when going from a system of $N$ spins to $N+1$ spins.
Since the new spin is inserted at the right end, it is clear, for instance, that the subset $\{\uparrow\uparrow\}$ of the new $N+1$ system will emerge from the union of the sets $\{\uparrow\uparrow\}$ and $\{\uparrow\downarrow\}$ of the $N$-spin chain with an extra spin up $S_{N+1}=1$.
This case is shown schematically in Fig.~\ref{fig}~B.
Consequently, the new term $Z_{N+1}^{\uparrow\uparrow}$ will have two contributions, one involving   
$Z_{N}^{\uparrow\uparrow}$
and the other involving $Z_{N}^{\uparrow\downarrow}$.
The first contribution comes from adding a spin $+1$ at the site $N+1$ to the previous set $\{\uparrow\uparrow\}$.
The second contribution comes from adding a spin $+1$ at the site $N+1$ to the previous set $\{\uparrow\downarrow\}$.
\begin{eqnarray}
Z_{N+1}^{\uparrow\uparrow} = c_1Z_{N}^{\uparrow\uparrow} + c_2Z_{N}^{\uparrow\downarrow},
\end{eqnarray}
where $c_1$ and $c_2$ are proper modifications to the Boltzmann weights.

The factors can be written down very simply. For example, in the first term, a new spin up is added, which contributes a factor $\exp(\beta h)$ from interactions with the field, two new satisfied bonds are created and one satisfied bond is eliminated, this contributes to a factor $\exp(\beta J)$. Putting the magnetic and bond contributions together, $c_1=\exp(\beta(h+J))$. Similarly, $c_2=\exp(\beta(h+J))$ as well. The whole transformation can be simply enumerated, introducing for simplicity a vector partition function
\begin{eqnarray}
\vec{z}_N=
\begin{pmatrix}
Z_N^{\uparrow \uparrow}, Z_N^{\uparrow \downarrow}, Z_N^{\downarrow \uparrow}, Z_N^{\downarrow \downarrow}
\end{pmatrix}
^T,
\end{eqnarray}
in such a way that the partition function of the system can be expressed as a summation reduction $Z_N = \sum (\vec{z}_N)$. 
And the recurrence relation can be expressed concisely in the form
\begin{eqnarray}
\vec{z}_{N+1} &=& M \vec{z}_N,
\label{matrix0}
\end{eqnarray}
where the recurrence matrix $M$ is
\begin{eqnarray}
M=
\begin{pmatrix}
e^{\beta(J+h)} &e^{\beta(J+h)} &0 &0 \\
e^{-\beta (3J+h)} &e^{\beta(J-h)} &0 &0 \\
0 &0 &e^{\beta(J+h)} &e^{-\beta (3J-h)} \\
0 &0 &e^{\beta(J-h)} &e^{\beta(J-h)}
\end{pmatrix}
\end{eqnarray}
Note that in spite of the matrix is $4\times4$ compared with the transfer matrix, it is block diagonal. Indeed we will see in the following the partition function, upon summation reduction, takes exactly the same form as that of the transfer matrix method. And most importantly for the learning context, the derivation of this recurrent relation is absolutely intuitive, it comes just from considering the energetic effect of inserting a new spin in the chain.
This is the most important step of the method, the rest is straightforward algebraic calculations.

Next the job is to find the vector partition function for a small system to start the induction e.g. a two-spin system. It is natural to start from $N=2$, however, one can check that it is also possible to start from $N=1$, i.e., a single spin on a ring which loops over and interact with itself. In this case we have
\begin{eqnarray}
\vec{z}_1=
\begin{pmatrix}
\exp(\beta(J+h)),0, 0, \exp(\beta(J-h))
\end{pmatrix}
^T.
\end{eqnarray}

Therefore, we readily have $\vec{z}_N = M^{N-1} \vec{z}_1$. 
As $M$ is a block diagonal matrix, let us call $M_{11}$ and $M_{22}$ to the first and last non-zero blocks respectively, we then write
\begin{eqnarray}
\vec{z}_N &=&
\begin{pmatrix}
M_{11}^{N-1} &0 \\
0 &M_{22}^{N-1}
\end{pmatrix}
\vec{z}_1 \\
 &=&
\begin{pmatrix}
M_{11}^{N-1} \vec{z}_1(1:2) \\
M_{22}^{N-1} \vec{z}_1(3:4)
\end{pmatrix},
\end{eqnarray}
where $\vec{u}(m:n)$ is the vector formed by the elements from $m$-th to $n$-th of the vector $\vec{u}$. Naturally,
\begin{eqnarray}
Z_N &=& \sum(\vec{z}_N) = \sum(\vec{z}_N(1:2)) + \sum(\vec{z}_N(3:4)).
\end{eqnarray}
Note that we are going in this step from a $4\times4$ matrix towards a $2\times2$ matrix upon the summation reduction, the same matrix size as the transfer matrix.

Now we introduce a lemma which is easy to prove:

\textit{Lemma:} If $A$ is a $2\times2$ matrix and $b$ is a $2 \times 1$ vector, then $\sum(Ab) = Tr(A \times [b,b])$.

Therefore,
\begin{eqnarray}
Z_N &=& Tr \left(M_{11}^{N-1} [\vec{z}_1(1:2),\vec{z}_1(1:2)]\right) + Tr \left(M_{22}^{N-1} [\vec{z}_1(3:4),\vec{z}_1(3:4)]\right) \\
\label{ZN}
&=& Tr \left(M_{11}^{N-1} [\vec{z}_1(1:2),\vec{z}_1(1:2)] + M_{22}^{N-1} [\vec{z}_1(3:4),\vec{z}_1(3:4)]\right).
\end{eqnarray}

Both $M_{11}$ and $M_{22}$ can be easily symmetrized using similarity transformations
\begin{eqnarray}
M_{11} &=&
\begin{pmatrix}
e^{\beta (J + h/2)} &0 \\
0 &e^{-\beta (J + h/2)} %aaaaaaaaaaaaaaaaaaaaaaaa
\end{pmatrix}
\begin{pmatrix}
e^{\beta (J+h)} &e^{-\beta J} \\
e^{-\beta J} &e^{\beta (J-h)}
\end{pmatrix}
\begin{pmatrix}
e^{-\beta (J + h/2)} &0 \\
0 &e^{\beta (J + h/2)}
\end{pmatrix} \\
M_{22} &=&
\begin{pmatrix}
e^{-\beta (J - h/2)} &0 \\
0 &e^{\beta (J - h/2)} %aaaaaaaaaaaaaaaaaaaaaaaa
\end{pmatrix}
\begin{pmatrix}
e^{\beta (J+h)} &e^{-\beta J} \\
e^{-\beta J} &e^{\beta (J-h)}
\end{pmatrix}
\begin{pmatrix}
e^{\beta (J - h/2)} &0 \\
0 &e^{-\beta (J - h/2)}
\end{pmatrix}
\end{eqnarray}

Plugging these two equations to $Z_N$, Eq.~\ref{ZN} remarkably simplifies to
\begin{eqnarray}
Z_N = Tr
\begin{pmatrix}
e^{\beta (J+h)} &e^{-\beta J} \\
e^{-\beta J} &e^{\beta (J-h)}
\end{pmatrix}^N,
\end{eqnarray}
which is identical to the transfer matrix result in Eq.~(\ref{tmres}).

Finally, we mention that the method can also be used in a purely numerical manner directly from the recurrence relation. This is computationally appealing because the setup is straightforward, and it can be easily applied to e.g. free boundary conditions. While this may seem unnecessary for our example, the method can be readily generalized to disordered systems such as random ferromagnets~\cite{theodorou82,avgin96} and even spin glasses \cite{EA,Machta:PA} with competing interactions.

%\section{The crossover length}
%Can we compute the correlation length critical exponent $\nu$ from the following simple crossover length? Start from a spin up, and then try to add spins in a heat-bath way, then we look at the number of spins added with spin up until hitting a spin down. The number of spins added has support $n=0, 1, 2, ...$ with a geometrical distribution
%$P_n = p^n(1-p)$, where $p=\exp(\beta J)/[\exp(\beta J) + \exp(-\beta J)]$. It is easy to compute the mean of this random variable and we get the crossover length $L_C = \langle n \rangle = p/(1-p) = \exp(2 \beta J)$.
%
%While the length grows and diverges at $T=0$ as expected, it is embarrassing that only $\ln(L_C) = 2J/(k_B T)$, we cannot get the the critical exponent $\nu$, and it appears the crossover length is very different from and exponentially larger than the correlation length. It is not very clear why this is so. It may have something to do with the stability of the chains and the length $L_C$ is a much dramatic overestimation of the true correlation length.

\section{Conclusions}

In this work, we have presented a mathematical induction approach to the solution of the 1D Ising model, complementing the more traditional transfer matrix method.
The analytical procedure is rather intuitive, physically clear, and computationally friendly, which makes the formalism very suited to be used in introductory statistical mechanics courses.
Remarkably, the algebraic derivation following this path connects with the formalism of the transfer matrix method. This fully equivalence is an extra benefit since it provides a physical interpretation of the transfer matrix, that appears naturally from considering the recurrent summation of the energetic cost of inserting a new spin in the chain.

\acknowledgments
W.W.~gratefully acknowledges support from the Swedish Research Council Grant No.~642-2013-7837 and Goran Gustafsson Foundation for Research in
Natural Sciences and Medicine.
%Computations were performed on resources provided by the Swedish National Infrastructure for Computing (SNIC) on the Triolith cluster at NSC and the Kebnekaise cluster at HPC2N.
%The computations were performed on resources
%provided by the Swedish National Infrastructure for Computing (SNIC)
%at the National Supercomputer Centre (NSC) and the High Performance Computing Center North (HPC2N).

\bibliography{ising}

\begin{thebibliography}{28}
\expandafter\ifx\csname natexlab\endcsname\relax\def\natexlab#1{#1}\fi
\expandafter\ifx\csname bibnamefont\endcsname\relax
  \def\bibnamefont#1{#1}\fi
\expandafter\ifx\csname bibfnamefont\endcsname\relax
  \def\bibfnamefont#1{#1}\fi
\expandafter\ifx\csname citenamefont\endcsname\relax
  \def\citenamefont#1{#1}\fi
\expandafter\ifx\csname url\endcsname\relax
  \def\url#1{\texttt{#1}}\fi
\expandafter\ifx\csname urlprefix\endcsname\relax\def\urlprefix{URL }\fi
\providecommand{\bibinfo}[2]{#2}
\providecommand{\eprint}[2][]{\url{#2}}

\bibitem[{\citenamefont{Ising}(1925)}]{ising25}
\bibinfo{author}{\bibfnamefont{E.}~\bibnamefont{Ising}},
  \emph{\bibinfo{title}{{Beitrag zur theorie des ferromagnetismus}}},
  \bibinfo{journal}{Zeit. Phys.} \textbf{\bibinfo{volume}{31}},
  \bibinfo{pages}{253} (\bibinfo{year}{1925}).

\bibitem[{\citenamefont{Brush}(1967)}]{brush67}
\bibinfo{author}{\bibfnamefont{S.~G.} \bibnamefont{Brush}},
  \emph{\bibinfo{title}{{History of the Lenz-Ising model}}},
  \bibinfo{journal}{Rev. Mod. Phys.} \textbf{\bibinfo{volume}{39}},
  \bibinfo{pages}{883} (\bibinfo{year}{1967}).

\bibitem[{\citenamefont{Taroni}(2015)}]{Taroni15}
\bibinfo{author}{\bibfnamefont{A.}~\bibnamefont{Taroni}},
  \emph{\bibinfo{title}{{Statistical physics: 90 years of the Ising model}}},
  \bibinfo{journal}{Nature Physics} \textbf{\bibinfo{volume}{11}},
  \bibinfo{pages}{997} (\bibinfo{year}{2015}).

\bibitem[{\citenamefont{Onsager}(1944)}]{onsager44}
\bibinfo{author}{\bibfnamefont{L.}~\bibnamefont{Onsager}},
  \emph{\bibinfo{title}{{Crystal statistics: I. A two-dimensional model with an
  order-disorder transition}}}, \bibinfo{journal}{Phys. Rev.}
  \textbf{\bibinfo{volume}{65}}, \bibinfo{pages}{117} (\bibinfo{year}{1944}).

\bibitem[{\citenamefont{Kac and Ward}(1952)}]{Kac52}
\bibinfo{author}{\bibfnamefont{M.}~\bibnamefont{Kac}} \bibnamefont{and}
  \bibinfo{author}{\bibfnamefont{J.~C.} \bibnamefont{Ward}},
  \emph{\bibinfo{title}{{A combinatorial solution of the two-dimensional Ising
  model}}}, \bibinfo{journal}{Phys. Rev.} \textbf{\bibinfo{volume}{88}},
  \bibinfo{pages}{1332} (\bibinfo{year}{1952}).

\bibitem[{\citenamefont{Kan\^{o} and Naya}(1953)}]{Kano53}
\bibinfo{author}{\bibfnamefont{K.}~\bibnamefont{Kan\^{o}}} \bibnamefont{and}
  \bibinfo{author}{\bibfnamefont{S.}~\bibnamefont{Naya}},
  \emph{\bibinfo{title}{{Antiferromagnetism. The Kagom\'{e} Ising Net}}},
  \bibinfo{journal}{Progress of Theoretical Physics}
  \textbf{\bibinfo{volume}{10}}, \bibinfo{pages}{158} (\bibinfo{year}{1953}).

\bibitem[{\citenamefont{Domb}(1960)}]{Domb60}
\bibinfo{author}{\bibfnamefont{C.}~\bibnamefont{Domb}},
  \emph{\bibinfo{title}{On the theory of cooperative phenomena in crystals}},
  \bibinfo{journal}{Advances in Physics} \textbf{\bibinfo{volume}{9}},
  \bibinfo{pages}{149} (\bibinfo{year}{1960}).

\bibitem[{\citenamefont{J.Kofinger and Dellago}(2010)}]{kofinger10}
\bibinfo{author}{\bibnamefont{J.Kofinger}} \bibnamefont{and}
  \bibinfo{author}{\bibfnamefont{C.}~\bibnamefont{Dellago}},
  \emph{\bibinfo{title}{{Single-file water as a one-dimensional Ising model}}},
  \bibinfo{journal}{New J. Phys.} \textbf{\bibinfo{volume}{12}},
  \bibinfo{pages}{09304} (\bibinfo{year}{2010}).

\bibitem[{\citenamefont{Sarkanych et~al.}(2018)\citenamefont{Sarkanych,
  Holovatch, and Kenna}}]{sarkanych18}
\bibinfo{author}{\bibfnamefont{P.}~\bibnamefont{Sarkanych}},
  \bibinfo{author}{\bibfnamefont{Y.}~\bibnamefont{Holovatch}},
  \bibnamefont{and} \bibinfo{author}{\bibfnamefont{R.}~\bibnamefont{Kenna}},
  \emph{\bibinfo{title}{{Classical phase transitions in a one-dimensional
  short-range spin model}}}, \bibinfo{journal}{J. Phys. A: Math. Theor.}
  \textbf{\bibinfo{volume}{51}}, \bibinfo{pages}{505001}
  (\bibinfo{year}{2018}).

\bibitem[{\citenamefont{Stre\ifmmode~\check{c}\else \v{c}\fi{}ka
  et~al.}(2008)\citenamefont{Stre\ifmmode~\check{c}\else \v{c}\fi{}ka,
  \ifmmode~\check{C}\else \v{C}\fi{}anov\'a, Ja\ifmmode \check{s}\else
  \v{s}\fi{}\ifmmode~\check{c}\else \v{c}\fi{}ur, and Hagiwara}}]{HM08}
\bibinfo{author}{\bibfnamefont{J.}~\bibnamefont{Stre\ifmmode~\check{c}\else
  \v{c}\fi{}ka}},
  \bibinfo{author}{\bibfnamefont{L.}~\bibnamefont{\ifmmode~\check{C}\else
  \v{C}\fi{}anov\'a}},
  \bibinfo{author}{\bibfnamefont{M.}~\bibnamefont{Ja\ifmmode \check{s}\else
  \v{s}\fi{}\ifmmode~\check{c}\else \v{c}\fi{}ur}}, \bibnamefont{and}
  \bibinfo{author}{\bibfnamefont{M.}~\bibnamefont{Hagiwara}},
  \emph{\bibinfo{title}{{Exact solution of the geometrically frustrated
  spin-$\frac{1}{2}$ Ising-Heisenberg model on the triangulated kagome
  (triangles-in-triangles) lattice}}}, \bibinfo{journal}{Phys. Rev. B}
  \textbf{\bibinfo{volume}{78}}, \bibinfo{pages}{024427}
  (\bibinfo{year}{2008}).

\bibitem[{\citenamefont{Bach et~al.}(2019)\citenamefont{Bach, Nguyen, and
  Bach}}]{bach19}
\bibinfo{author}{\bibfnamefont{C.~T.} \bibnamefont{Bach}},
  \bibinfo{author}{\bibfnamefont{N.~T.} \bibnamefont{Nguyen}},
  \bibnamefont{and} \bibinfo{author}{\bibfnamefont{G.~H.} \bibnamefont{Bach}},
  \emph{\bibinfo{title}{{Thermodynamic properties of ferroics described by the
  transverse Ising model and their applications for CoNb$_2$O$_6$}}},
  \bibinfo{journal}{J. Mag. Mag. Mater.} \textbf{\bibinfo{volume}{483}},
  \bibinfo{pages}{136} (\bibinfo{year}{2019}).

\bibitem[{\citenamefont{Baxter}(1982)}]{Baxter82}
\bibinfo{author}{\bibfnamefont{R.~J.} \bibnamefont{Baxter}},
  \emph{\bibinfo{title}{{Exactly Solved Models in Statistical Mechanics}}}
  (\bibinfo{publisher}{New York: Academic}, \bibinfo{year}{1982}).

\bibitem[{\citenamefont{Seth}(2017)}]{Seth17}
\bibinfo{author}{\bibfnamefont{S.}~\bibnamefont{Seth}},
  \emph{\bibinfo{title}{{Combinatorial approach to exactly solve the 1D Ising
  model}}}, \bibinfo{journal}{Eur. J. Phys.} \textbf{\bibinfo{volume}{38}},
  \bibinfo{pages}{015104} (\bibinfo{year}{2017}).

\bibitem[{\citenamefont{Moreno et~al.}(2005)\citenamefont{Moreno,
  S{\'a}nchez-L{\'o}pez, Ferreira, Davis, and Mateos}}]{moreno05}
\bibinfo{author}{\bibfnamefont{I.}~\bibnamefont{Moreno}},
  \bibinfo{author}{\bibfnamefont{M.~M.} \bibnamefont{S{\'a}nchez-L{\'o}pez}},
  \bibinfo{author}{\bibfnamefont{C.}~\bibnamefont{Ferreira}},
  \bibinfo{author}{\bibfnamefont{J.~A.} \bibnamefont{Davis}}, \bibnamefont{and}
  \bibinfo{author}{\bibfnamefont{F.}~\bibnamefont{Mateos}},
  \emph{\bibinfo{title}{{Teaching Fourier optics through ray matrices}}},
  \bibinfo{journal}{Eur. J. Phys.} \textbf{\bibinfo{volume}{26}},
  \bibinfo{pages}{261} (\bibinfo{year}{2005}).

\bibitem[{\citenamefont{Pujol and P{\'e}rez}(2007)}]{pujol07}
\bibinfo{author}{\bibfnamefont{O.}~\bibnamefont{Pujol}} \bibnamefont{and}
  \bibinfo{author}{\bibfnamefont{J.~P.} \bibnamefont{P{\'e}rez}},
  \emph{\bibinfo{title}{{A synthetic approach to the transfer matrix method in
  classical and quantum physics}}}, \bibinfo{journal}{Eur. J. Phys.}
  \textbf{\bibinfo{volume}{28}}, \bibinfo{pages}{679} (\bibinfo{year}{2007}).

\bibitem[{\citenamefont{Pujol et~al.}(2014)\citenamefont{Pujol, Carles, and
  P{\'e}rez}}]{pujol14}
\bibinfo{author}{\bibfnamefont{O.}~\bibnamefont{Pujol}},
  \bibinfo{author}{\bibfnamefont{R.}~\bibnamefont{Carles}}, \bibnamefont{and}
  \bibinfo{author}{\bibfnamefont{J.~P.} \bibnamefont{P{\'e}rez}},
  \emph{\bibinfo{title}{{Quantum propagation and confinement in 1D systems
  using the transfer-matrix method}}}, \bibinfo{journal}{Eur. J. Phys.}
  \textbf{\bibinfo{volume}{35}}, \bibinfo{pages}{035025}
  (\bibinfo{year}{2014}).

\bibitem[{\citenamefont{Zad and Ananikian}(2017)}]{Zad17}
\bibinfo{author}{\bibfnamefont{H.~A.} \bibnamefont{Zad}} \bibnamefont{and}
  \bibinfo{author}{\bibfnamefont{N.}~\bibnamefont{Ananikian}},
  \emph{\bibinfo{title}{{Phase transitions and thermal entanglement of the
  distorted Ising{\textendash}Heisenberg spin chain: topology of multiple-spin
  exchange interactions in spin ladders}}}, \bibinfo{journal}{Journal of
  Physics: Condensed Matter} \textbf{\bibinfo{volume}{29}},
  \bibinfo{pages}{455402} (\bibinfo{year}{2017}).

\bibitem[{\citenamefont{Torrico et~al.}(2018)\citenamefont{Torrico,
  Stre\u{c}ka, Hagiwara, Rojas, de~Souza, Han, Honda, and Lyra}}]{Torrico18}
\bibinfo{author}{\bibfnamefont{J.}~\bibnamefont{Torrico}},
  \bibinfo{author}{\bibfnamefont{J.}~\bibnamefont{Stre\u{c}ka}},
  \bibinfo{author}{\bibfnamefont{M.}~\bibnamefont{Hagiwara}},
  \bibinfo{author}{\bibfnamefont{O.}~\bibnamefont{Rojas}},
  \bibinfo{author}{\bibfnamefont{S.}~\bibnamefont{de~Souza}},
  \bibinfo{author}{\bibfnamefont{Y.}~\bibnamefont{Han}},
  \bibinfo{author}{\bibfnamefont{Z.}~\bibnamefont{Honda}}, \bibnamefont{and}
  \bibinfo{author}{\bibfnamefont{M.}~\bibnamefont{Lyra}},
  \emph{\bibinfo{title}{{Heterobimetallic Dy-Cu coordination compound as a
  classical-quantum ferrimagnetic chain of regularly alternating Ising and
  Heisenberg spins}}}, \bibinfo{journal}{Journal of Magnetism and Magnetic
  Materials} \textbf{\bibinfo{volume}{460}}, \bibinfo{pages}{368 }
  (\bibinfo{year}{2018}).

\bibitem[{\citenamefont{Kassan-Ogly}(2001)}]{kassan01}
\bibinfo{author}{\bibfnamefont{F.~A.} \bibnamefont{Kassan-Ogly}},
  \emph{\bibinfo{title}{{One-dimensional ising model with
  next-nearest-neighbour interaction in magnetic field}}},
  \bibinfo{journal}{Phase Transitions} \textbf{\bibinfo{volume}{74}},
  \bibinfo{pages}{353} (\bibinfo{year}{2001}).

\bibitem[{\citenamefont{Tenga et~al.}(2002)\citenamefont{Tenga, Chenb, Fuc,
  Tangc, Tuc, Chenc, and Tang}}]{baohua02}
\bibinfo{author}{\bibfnamefont{B.}~\bibnamefont{Tenga}},
  \bibinfo{author}{\bibfnamefont{Y.}~\bibnamefont{Chenb}},
  \bibinfo{author}{\bibfnamefont{H.}~\bibnamefont{Fuc}},
  \bibinfo{author}{\bibfnamefont{Y.}~\bibnamefont{Tangc}},
  \bibinfo{author}{\bibfnamefont{M.}~\bibnamefont{Tuc}},
  \bibinfo{author}{\bibfnamefont{Y.}~\bibnamefont{Chenc}}, \bibnamefont{and}
  \bibinfo{author}{\bibfnamefont{J.}~\bibnamefont{Tang}},
  \emph{\bibinfo{title}{{Comparison between the nearest and the next-nearest
  neighbor site -- spin interactions in the Ising model}}},
  \bibinfo{journal}{Solid State Comm.} \textbf{\bibinfo{volume}{124}},
  \bibinfo{pages}{347} (\bibinfo{year}{2002}).

\bibitem[{\citenamefont{Shrock and Tsai}(1997)}]{shrock97}
\bibinfo{author}{\bibfnamefont{R.}~\bibnamefont{Shrock}} \bibnamefont{and}
  \bibinfo{author}{\bibfnamefont{S.-H.} \bibnamefont{Tsai}},
  \emph{\bibinfo{title}{{Ground-State Entropy of Potts Antiferromagnets:
  Bounds, Series, and Monte Carlo Measurements}}}, \bibinfo{journal}{Phys. Rev.
  E} \textbf{\bibinfo{volume}{56}}, \bibinfo{pages}{2733}
  (\bibinfo{year}{1997}).

\bibitem[{\citenamefont{Shrock and Xu}(2010)}]{shrock10}
\bibinfo{author}{\bibfnamefont{R.}~\bibnamefont{Shrock}} \bibnamefont{and}
  \bibinfo{author}{\bibfnamefont{Y.}~\bibnamefont{Xu}},
  \emph{\bibinfo{title}{{Exact Results on Potts Model Partition Functions in a
  Generalized External Field and Weighted-Set Graph Colorings}}},
  \bibinfo{journal}{J. Stat. Phys.} \textbf{\bibinfo{volume}{141}},
  \bibinfo{pages}{909} (\bibinfo{year}{2010}).

\bibitem[{\citenamefont{Yurishchev}(2001)}]{yurishchev01}
\bibinfo{author}{\bibfnamefont{M.~A.} \bibnamefont{Yurishchev}},
  \emph{\bibinfo{title}{{Double potts chain and exact results for some
  two-dimensional spin models}}}, \bibinfo{journal}{J. Exp. Theor. Phys.}
  \textbf{\bibinfo{volume}{93}}, \bibinfo{pages}{1113} (\bibinfo{year}{2001}).

\bibitem[{\citenamefont{Guidi et~al.}(2015)\citenamefont{Guidi, Gillon, Mason,
  Garlatti, Carretta, Santini, Stunault, Caciuffo, van Slageren, Klemke
  et~al.}}]{guidi15}
\bibinfo{author}{\bibfnamefont{T.}~\bibnamefont{Guidi}},
  \bibinfo{author}{\bibfnamefont{B.}~\bibnamefont{Gillon}},
  \bibinfo{author}{\bibfnamefont{S.~A.} \bibnamefont{Mason}},
  \bibinfo{author}{\bibfnamefont{E.}~\bibnamefont{Garlatti}},
  \bibinfo{author}{\bibfnamefont{S.}~\bibnamefont{Carretta}},
  \bibinfo{author}{\bibfnamefont{P.}~\bibnamefont{Santini}},
  \bibinfo{author}{\bibfnamefont{A.}~\bibnamefont{Stunault}},
  \bibinfo{author}{\bibfnamefont{R.}~\bibnamefont{Caciuffo}},
  \bibinfo{author}{\bibfnamefont{J.}~\bibnamefont{van Slageren}},
  \bibinfo{author}{\bibfnamefont{B.}~\bibnamefont{Klemke}},
  \bibnamefont{et~al.}, \emph{\bibinfo{title}{{Direct observation of finite
  size effects in chains of antiferromagnetically coupled spins}}},
  \bibinfo{journal}{Nat. Comm.} \textbf{\bibinfo{volume}{6}},
  \bibinfo{pages}{7061} (\bibinfo{year}{2015}).

\bibitem[{\citenamefont{Theodorou}(1982)}]{theodorou82}
\bibinfo{author}{\bibfnamefont{G.}~\bibnamefont{Theodorou}},
  \emph{\bibinfo{title}{{Spin waves in random one-dimensional ferromagnets}}},
  \bibinfo{journal}{J. Phys. C: Solid State Phys.}
  \textbf{\bibinfo{volume}{15}}, \bibinfo{pages}{L1315} (\bibinfo{year}{1982}).

\bibitem[{\citenamefont{Avgin}(1996)}]{avgin96}
\bibinfo{author}{\bibfnamefont{I.}~\bibnamefont{Avgin}},
  \emph{\bibinfo{title}{{A ferromagnetic chain in a random weak field}}},
  \bibinfo{journal}{J. Phys.: Condens. Matter} \textbf{\bibinfo{volume}{8}},
  \bibinfo{pages}{8379} (\bibinfo{year}{1996}).

\bibitem[{\citenamefont{Edwards and Anderson}(1975)}]{EA}
\bibinfo{author}{\bibfnamefont{S.~F.} \bibnamefont{Edwards}} \bibnamefont{and}
  \bibinfo{author}{\bibfnamefont{P.~W.} \bibnamefont{Anderson}},
  \emph{\bibinfo{title}{{Theory of spin glasses}}}, \bibinfo{journal}{J. Phys.
  F: Met. Phys.} \textbf{\bibinfo{volume}{5}}, \bibinfo{pages}{965}
  (\bibinfo{year}{1975}).

\bibitem[{\citenamefont{Machta}(2010)}]{Machta:PA}
\bibinfo{author}{\bibfnamefont{J.}~\bibnamefont{Machta}},
  \emph{\bibinfo{title}{{Population annealing with weighted averages: A {M}onte
  {C}arlo method for rough free-energy landscapes}}}, \bibinfo{journal}{Phys.
  Rev. E} \textbf{\bibinfo{volume}{82}}, \bibinfo{pages}{026704}
  (\bibinfo{year}{2010}).

\end{thebibliography}

\end{document}